\documentclass[fleqn,usenatbib]{mnras}

\usepackage{newtxtext,newtxmath}
\usepackage{tablefootnote}

\usepackage[T1]{fontenc}
\usepackage{longtable}
\DeclareRobustCommand{\VAN}[3]{#2}
\let\VANthebibliography\thebibliography
\def\thebibliography{\DeclareRobustCommand{\VAN}[3]{##3}\VANthebibliography}

\usepackage{xcolor}
\usepackage{graphicx}	
\usepackage{amsmath}	

\def\redpen#1{\textcolor{black}{#1}}

\title[Stealth Binaries]{Do anomalously-dense hot Jupiters orbit stealth binary stars?}

\author[Goswamy, Collier Cameron \& Wilson]{
Tanvi Goswamy$^{1}$,
Andrew Collier Cameron$^{1}$\thanks{acc4@st-andrews.ac.uk},
Thomas G. Wilson$^{1}$
\\
$^{1}$Centre for Exoplanet Science, SUPA School of Physics and Astronomy, University of St Andrews, St Andrews, Fife KY16 9SS, UK
}

\date{Accepted 2024 September 02. Received 2024 August 26; in original form 2024 July 03}

\pubyear{2024}

\begin{document}
\label{firstpage}
\pagerange{\pageref{firstpage}--\pageref{lastpage}}
\maketitle

\begin{abstract}
The Wide Angle Search for Planets (WASP) survey used transit photometry to discover nearly 200 gas-giant exoplanets and derive their planetary and stellar parameters. 
Reliable determination of the planetary density depends on accurate measurement of the planet's radius, obtained from the transit depth and photodynamical determination of the stellar radius. The stellar density, and hence the stellar radius are typically determined in a model-independent way from the star's reflex orbital acceleration and the transit profile. Additional flux coming from the system due to a bright, undetected stellar binary companion can, however, potentially dilute
the transit curve and radial velocity signal, leading to under-estimation of the planet's mass and radius, and to overestimation of the planet’s 
density. 
In this study, we cross-check the published radii of all the WASP planet host stars, determined from their transit profiles and radial-velocity curves, against radiometric measurements of stellar radii derived from their angular diameters (via the Infrared Flux method) and trigonometric parallaxes. 
We identify eight systems showing radiometric stellar radii significantly greater than their published photodynamical values: WASPs 20, 85, 86, 103, 105, 129, 144 and 171. We
investigate these systems in more detail to establish plausible ranges of angular and radial-velocity separations within which such ``stealth binaries'' could evade detection,
and deduce their likely
orbital periods, mass ratios, and flux ratios. 
After accounting for
the dilution of transit depth and radial velocity amplitude, we
find that on average, 
the planetary densities for the identified stealth binary systems 
should be reduced
by a factor of 1.3.
\end{abstract}

\begin{keywords}
\redpen{
stars: planetary systems -- 
stars: binaries: general --
planets and satellites: fundamental parameters planets --
planets and satellites: gaseous planets --
techniques: photometric --
techniques: spectroscopic
}
\end{keywords}



\section{Introduction}

The Wide-Angle Search for Planets (WASP) project \citep{2006PASP..118.1407P} 
has published the discoveries of \redpen{over 178 transiting gas-giant exoplanets in 
close orbits about their host stars \citep{2011MNRAS.417.2166S}. }

The validation and characterisation of a WASP planet candidate involves
photodynamical analysis of the transit light curve to establish the \redpen{planet/star radius 
ratio and the stellar density from the transit depth and duration respectively \citep{2010MNRAS.408.1689S}.}
Since most WASP host stars have masses close to solar, the inverse cube root 
of the stellar density (in solar units) provides an approximate estimate of 
the stellar radius. Follow-up spectroscopy yields the stellar effective 
temperature, surface gravity and metallicity. This spectroscopic 
characterisation allows the stellar mass to be estimated. Radial-velocity 
observations of the host star's orbit then determine the planetary mass.

Since the release of the {\em Gaia} DR2 and 
DR3
catalogues 
\citep{2021A&A...649A...1G,2023A&A...674A...1G},
the availability of precise 
parallaxes has made it possible to determine WASP host-star radii independently. The 
stellar angular diameter can be estimated from the effective temperature 
and an estimate  of the  bolometric flux received at Earth, via the Infrared Flux method ("IRFM") 
of \cite{{1977MNRAS.180..177B}}. The angular diameter and {\em Gaia} parallax 
together yield a direct geometrical estimate of the the stellar radius.
The radii determined via this method can be compared directly with the radii inferred from 
photodynamical fits to their planets' transit profiles.

The stellar radius estimates obtained via these two methods should agree 
unless the light of the host star is diluted significantly by a stellar 
binary companion. In such cases, the additional flux dilutes both the 
transit depth and the radial velocity amplitude. 
This in turn leads to an overestimation of the planet’s density and an 
underestimation of its radius and mass. Stellar binaries are usually 
detectable if the orbit is small enough that the Doppler-shifted 
spectral lines of the two stars are resolvable; or wide enough 
for the binary to be resolved through direct imaging. We coin the term 
"stealth binaries" to characterise systems whose orbital separations 
lie in the range in which they cannot be resolved by either method.

\cite{2015ApJ...805...16C} have carried out a similar study, discussing the effects of undetected multiplicity on planetary radii for Kepler Objects of Interest (KOIs). For each KOI, they found the best-fit Dartmouth isochrone, and considered all stars along the isochrones that had absolute Kepler magnitudes fainter than the primaries as viable companions. Their derivation of the theoretical correction factor $X_{R}$ - by which the planetary radius would have been underestimated -  is similar to the equations derived in Sections 4.3 and 4.4 of this paper. They derived mean values of $X_{R}$ for all possible scenarios up to a multiplicity of 3. \cite{2018AJ....156..209P} developed this concept further, using the relationship between the mean stellar density and stellar effective temperature to identify which of the stellar components in eight marginally-resolved multi-star systems were the hosts of transiting planets discovered with Kepler/K2.

In this paper we compare the stellar radii obtained from the 
spectral-energy distributions and $Gaia$ parallaxes of a large sample of  WASP planet-host stars, against those calculated from transit fitting and spectroscopic characterisation. In the majority of cases the host stars are not known a priori to be binaries. However, if previously-undetected binaries are present in the WASP catalogue, \redpen{the photodynamical host-star radii derived from the stellar density via the transit duration will be smaller than those determined from the stellar angular diameter and parallax. In cases where the discrepancy is significant,
their planets' bulk densities will have been over-estimated. }

\redpen{Our methods are discussed in detail in Section 2. In Section 3, we present the comparison of results from both methods. }
In Section 4.1, we examine these systems in more 
detail to predict the limits on angular separation and 
the limiting difference in radial velocities of the two stars 
that would allow the secondary star to remain undetected. 
These limits can be used to classify these systems 
as “stealth binaries” and estimate the plausible range of 
orbital periods of the stellar binaries. Section 4.2 discusses 
WASP-85AB – a known stellar binary which we 
use to verify some of our methods. In Section 4.3, 
we estimate the most probable flux ratio and mass 
ratio for each stealth binary system. The factor by 
which the radii from both methods differs gives us 
information about how much additional flux is 
being received from the system – and thus an 
estimate on the luminosity ratio of the two stars in 
the binary. We then use the \redpen{evolutionary tracks and isochrone tables of 
\cite{2012MNRAS.427..127B}} to estimate the binary mass ratios 
from the luminosity ratios. Finally, in Section 4.4, 
we assess the effect of contamination of 
observations on the derived system parameters, and 
hence recalculate them after having accounted for 
the secondary star. Corrections to parameters for some WASP planets based on similar studies have 
previously been made by \cite{2016ApJ...833L..19E}, 
\cite{2020A&A...635A..74S}, and \cite{2018MNRAS.474.2334D}; 
which will be discussed in Section 4.5. We conclude the study and 
suggest follow-up observations in Section 5.

\section{Methods}

We start by comparing the published radii of a sample of
178 host stars from the WASP survey, obtained by photodynamical 
modelling of their transit profiles, with radii estimated 
from their angular diameters and parallaxes. 

Photodynamical modelling of planetary transits is carried out routinely 
as part of the discovery process leading to the announcement of a new planet.
\cite{2009IAUS..253...99W} reviewed the 
planetary and stellar parameters that can be measured from precise 
photometry of exoplanet transits combined with radial-velocity follow-up.
The transit duration $(T)$ gives us an estimate of the stellar 
density as per Eq. \ref{eq:transit_dur} \textcolor{black}{for central transits}, and the transit depth ($\delta$) gives 
us the ratio of the planetary radius $(R_{p})$ to the stellar 
radius $(R_{s})$, as per the relation $\sqrt{\delta}$ = $R_{p}$/$R_{s}$. 

\begin{equation}
	\frac{T}{3\,{\rm h}} \approx  
	\left (\frac{P}{4\,{\rm days}}\right)^{1/3}*\left(\frac{\rho_{s}}	{\rho_{\odot}}\right)^{-1/3}
    \label{eq:transit_dur}
\end{equation}

Here, $P$ is the orbital period of the planet, and $\rho_{s}$ is 
the stellar density. Since most FGK main sequence 
stars have a mass close to unity, their radius $R_{s}$ can 
be estimated directly from the density. Dilution of 
the transit curve due to flux from a potential binary 
companion will not cause a change in the transit 
duration – which is why we can rely on the $R_{s}$
calculation from this method (hereon "$R_{\text{trans}}$") to compare
results from the IRFM. However, it will cause a 
decrease in $\delta$ and thus an underestimation of $R_{p}$.

$R_{s}$ can also be derived from the spectrum of the star 
using the IRFM, which gives us the star’s effective 
temperature ($T_{\text{eff}}$). This method was first proposed by 
\cite{1977MNRAS.180..177B}, after which various 
improved scales have been proposed. The basic idea 
of the IRFM is to compare the ratio between the 
bolometric flux ($f_{\text{bol}}$) and the flux in a given IR bandpass ($f_{\text{IR}}$) –
both received at the Earth’s atmosphere; to the ratio 
between the stellar surface bolometric flux ($\sigma T_{\text{eff}} ^4$) and the surface flux in the same IR bandpass ($F_{\text{IR, model}}$), which is determined theoretically using the stellar $T_{\text{eff}}$, stellar surface gravity log $g$, and [Fe/H] values 
from the star’s spectrum \citep{2010A&A...512A..54C}. 
This is shown in Eq. \ref{eq:T_eff}, where $T_{\text{eff}}$ is the only 
unknown quantity:
 \textcolor{black}{
\begin{equation}
   \frac{f_{\rm bol}}{f_{\rm IR}}=\frac{\sigma T_{\rm eff}^{4}}{F_{\rm IR, model}}
	\label{eq:T_eff}
\end{equation}
}
Using $T_{\text{eff}}$ along with the $f_{\text{bol}}$ received from the star 
and the parallax from {\em Gaia} 
DR3
\citep{2023A&A...674A...1G} extracted using VizieR\footnote{https://vizier.cds.unistra.fr/viz-bin/VizieR-3?-source=I/355/gaiadr3} \citep{2000A&AS..143...23O}, $R_{s}$ (hereon "$R_{\text{IRFM}}$") can be calculated using Eq. \ref{eq:F_bol}:

\begin{equation}
    f_{\text{bol}}=\sigma T_{\text{eff}} ^4\left ( \frac{R_{s}}{d} \right )^{2}
	\label{eq:F_bol}
\end{equation}
Here, $d$ is the distance to the star given by the inverse of the
parallax, and $\sigma$ is the Stefan-Boltzmann constant.

In this study we estimate the angular diameter 
by fitting the apparent magnitudes in eight optical/IR bandpasses:  
{\em Gaia} $BP$, $G$ and $RP$ \citep{2023A&A...674A...1G}; 
{\em 2MASS} $J$, $H$ and $Ks$ \citep{2006AJ....131.1163S}; 
and {\em WISE} $W1$ and $W2$ \citep{2010AJ....140.1868W}, with
synthetic photometry derived from the stellar model atmospheres of \cite{2003IAUS..210P.A20C}. \redpen{While other bandpasses can be used in addition to these, this set has two advantages: they are available for all the WASP host stars, and the angular-diameter values derived from them are independent of degeneracies between stellar effective temperature and interstellar reddening \citep{2020MNRAS.499..428S}.}
At the distances of typical WASP host stars,  the parallax and the angular diameter 
combine to yield the stellar radius to a precision
of order 1 to 2 percent. 
This is comparable to the precision achievable with asteroseismology (e.g. \citealt{2015MNRAS.452.2127S,2017ApJ...835..173S}).

Our aim is to compare the values of $R_{s}$ using both the 
methods above for all the WASP systems. The stars 
that have a larger $R_{s}$ from the IRFM than from photodynamical modelling (i.e., 
$R_{\text{IRFM}}>R_{\text{trans}}$) are our targets of interest.

\section{Results}

\begin{figure}
	\includegraphics[width=\columnwidth]{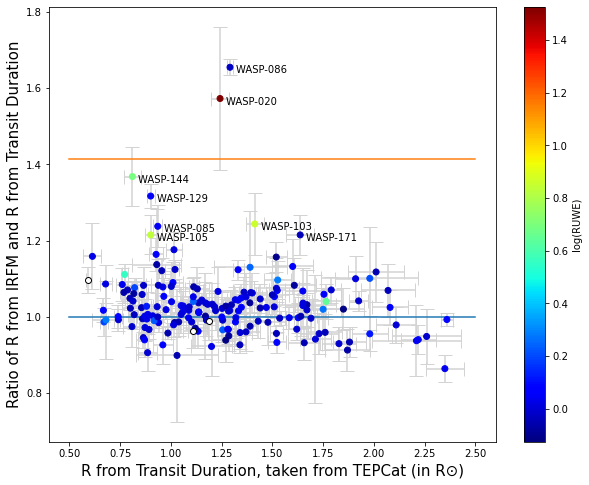}
    \caption{Ratio of radii from IRFM ($R_{\rm IRFM}$) and published photodynamical values taken from TEPCat ($R_{\rm trans}$) vs $R_{\rm trans}$, along with error bars. Most of the points lie along the equality line (light blue), while some of the stars (labelled in the graph) lie closer to, or even above, the $\sqrt{2}$ line (orange). The points are colour-coded based on their RUWE values taken from Gaia DR3, while the stars missing from the catalogue are in white.}
    \label{fig:TEPCat}
\end{figure}

We have used TEPCat\footnote{https://www.astro.keele.ac.uk/jkt/tepcat/allplanets-noerr.html} \citep{2011MNRAS.417.2166S} to extract all our data on the WASP systems. 
Because the discovery papers for the majority of these planets were published 
prior to the {\em Gaia} DR2 and DR3 data releases, the published $R_{s}$ values have in most cases been 
obtained using the transit duration method. We performed the IRFM calculations using Python 
routines\footnote{Routines used are in the file “GaiaIRFM\_EDR3.py”} developed by the authors, based on the {\sc astroquery} \citep{2021zndo...5804082G}, {\sc pysynphot} \citep{2013ascl.soft03023S} and {\sc pyphot} \citep{2022zndo...7016774F} packages, on all the systems to check the consistency with the 
published values. The inferred stellar radii are plotted against the values catalogued in TEPCat, in Fig. \ref{fig:TEPCat}. We have also extracted the renormalised unit weight error (RUWE) values for each star from Gaia DR3, which should be $\simeq$ 1 for single sources \citep{2023A&A...674A...1G}. 
The RUWE is similar in character to the reduced $\chi^2$ statistic; \cite{ziegler2020} note that values in excess of 1.4 are indicative of an extended or binary source \textcolor{black}{in {\em Gaia} DR2, while \citet{2022MNRAS.513.2437P} and \citet{2024A&A...688A...1C} suggest a lower threshold of order 1.25 for {\em Gaia} DR3.}
The points on the graph in Fig. \ref{fig:TEPCat} are colour-coded by their RUWE values to highlight outliers with high RUWE.

The radiometric stellar radii were derived from the angular diameter (obtained from the apparent bolometric flux and effective temperature derived from the synthetic photometry via eqs.~\ref{eq:T_eff} and \ref{eq:F_bol}) and the {\em Gaia} parallax. These were then compared with the photodynamical stellar radii $R_{\rm trans}$ obtained from TEPCat, which in the source publications had generally been computed under the assumption that no contaminating light was present.

We have identified 8 clear outliers in Fig.~\ref{fig:TEPCat} which lie significantly above the 
$R_{\rm trans}=R_{\rm IRFM}$ line, but near or below the $R_{\rm IRFM}=\sqrt{2}R_{\rm trans}$
line along which binaries with equal-luminosity components should lie. Six of these eight  WASP hosts (WASP-85, 103, 105, 129, 144, 171) have 
$R_{\rm{trans}} < R_{\rm{IRFM}} \lesssim \sqrt{2} R_{\rm{trans}}$, while WASPs 20 and 86 lie above the line. Half of these also have high RUWE values, as indicated by their colours on the graph, which provides further evidence of binarity \citep{2020MNRAS.496.1922B}. We note that WASP-86 \citep{2016arXiv160804225F}, has recently been identified as the same star as KELT-12 \citep{2017AJ....153..178S}. The apparent discrepancy in the stellar radii reported in these two discovery papers has recently been reconciled by \cite{SF21}.
This star has a long and shallow transit, causing the initial discrepancy in measurement of its properties in the two discovery papers. We will hence not include WASP-86 in our analysis. The revised radius can be found in \cite{SF21}. Out of the remaining \textcolor{black}{seven} stars, WASP-20 and WASP-85 are known binary systems. We will look at WASP-85 in more detail in a later section as we have sufficient information about its binary companion from the discovery paper itself. WASP-20 will be discussed in Section 4.5.

\section{Discussion}

Since we have identified the systems where we 
suspect the presence of a bright, unresolved secondary star diluting the 
transit curve, we can now begin our investigation of 
these systems in more detail. The next few sections 
discuss the various studies that we carried out on 
these 7 systems.

We first investigate the reasons why the companion 
star has not been detected yet.

\subsection{Radial velocity difference and angular 
separation}

\begin{figure}
	\includegraphics[width=\columnwidth]{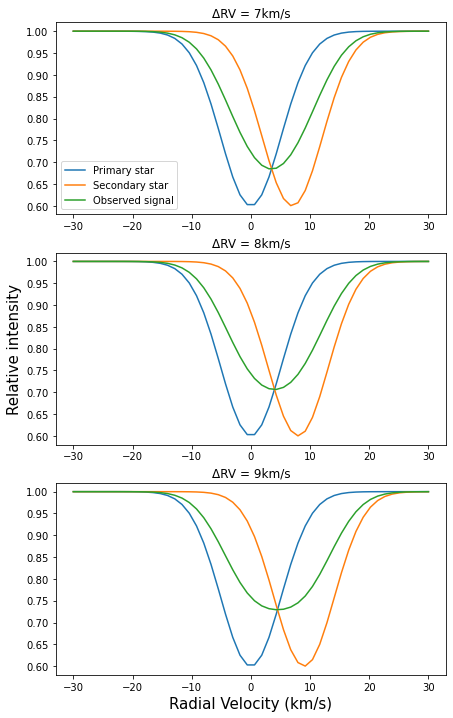}
    \caption {Synthetic cross-correlation functions of model binaries, comprising  
    pairs of gaussian profiles with radial-velocity separations of 7, 8 and 9~km~s$^{-1}$.
    A single broad peak is seen at separations up to 9~km~s$^{-1}$. At greater separations, 
    the composite profile becomes bimodal, and the binary nature becomes evident.
}
    \label{fig:RV}
\end{figure}

The transit depth can be diluted due to background stars that may not be bound to the primary star.
It is unlikely, however, that such a chance alignment would involve
two physically unrelated stars with indistinguishable radial velocities.
The absence of two resolvable spectra therefore allows us to assume that the 
two stars form a wide but bound pair, and that the dilution is 
caused by a secondary star that has a RV very close to that of the 
primary. For the two stars' spectra to be unresolved with a radial-velocity
spectrometer such as SOPHIE, CORALIE or HARPS, the Doppler shift between the two stars' spectral lines
would have to be close enough for their cross-correlation function to have a single 
peak. To estimate the detection threshold for such a binary, we approximated 
the CCFs of slowly-rotating solar-type stars as Gaussian profiles with the same width as 
the CCF of the Sun observed with the HARPS-N solar telescope feed \citep{2019MNRAS.487.1082C}, as shown in Fig. \ref{fig:RV}. 

For RV separations less than 
$\Delta$RV = 8 km/s, the combined signal appears 
like that of a single star, with no sign of the presence 
of a companion. Above that, the combined signal broadens 
and develops significant kurtosis, before separating into 
a recognisably bimodal profile.  We therefore  set our upper 
limit on for stealth binaries at $\Delta RV < 8$~km~s$^{-1}$.

It is also possible for stealth binaries to be mistaken for planet-host systems in the absence of high-resolution spectra, as discussed by \cite{2023AJ....165..266M}. They found that the primary false-positive scenario for astrometric exoplanet detections is an unresolved binary system with alike components and a small 
photocentric
orbit.

Another factor that can 
allow a contaminating secondary star to evade detection
during the RV follow-up is the 
angular separation ($\theta$) between the two stars. If $\theta$ is 
less than the fibre diameter of the instrument used, 
then the measured RV of the primary star will be 
contaminated. Spectroscopic follow-up for the 
WASP systems is usually done using the SOPHIE
\citep{2008SPIE.7014E..0JP} or CORALIE \citep{1996A&AS..119..373B,2000A&A...354...99Q}, instruments. These have fibre-aperture diameters of 3" and 2" respectively – which 
means any binaries with $\theta$ significantly less than the fibre diameter would 
not be detectable. However, the aperture size of the 
HARPS \citep{2003Msngr.114...20M} spectrograph is 1" and 
this instrument was used for follow-up on WASP-85, which is why we set our lower limit for 
detection at $\theta$ = 1".

Given the masses of the two stars and an estimate of the orbital period,
the physical separation $a$ between the two stars follows from Kepler’s 3rd law. 
Together with the parallax ($\hat{\pi}$) from Gaia, this yields the angular separation.
In solar units,
\begin{equation}
    \frac{\theta}{1\,{\rm arcsec}} =
    \left(\frac{\hat{\pi}}{1\,{\rm arcsec}}\right)
    \left(\frac{P}{1\,{\rm y}}\right)^{2/3}\left(\frac{M_{1}+M_{2}}{M_{\odot}}\right)^{1/3}.
    \label{eq:theta}
\end{equation}

The maximum radial-velocity separation $\Delta R V$, assuming a circular orbit
viewed edge-on, follows from energy conservation and Kepler’s 3rd law:
\begin{equation}
    \frac{\Delta RV}{v_{\oplus}} =
    \left(\frac{M_{1}+M_{2}}{M_{\odot}}\right)^{1/2}
    \left(\frac{a}{1\,{\rm au}}\right)^{-1/2}  =
    \left(\frac{M_{1}+M_{2}}{M_{\odot}}\right)^{1/3}
    \left(\frac{P}{1\,{\rm y}}\right)^{-1/3},
	\label{eq:deltaRV}
\end{equation}
scaling to the Earth's orbital velocity $v_{\oplus}\simeq 30$~km~s$^{-1}$.

\begin{figure}
	\includegraphics[width=\columnwidth]{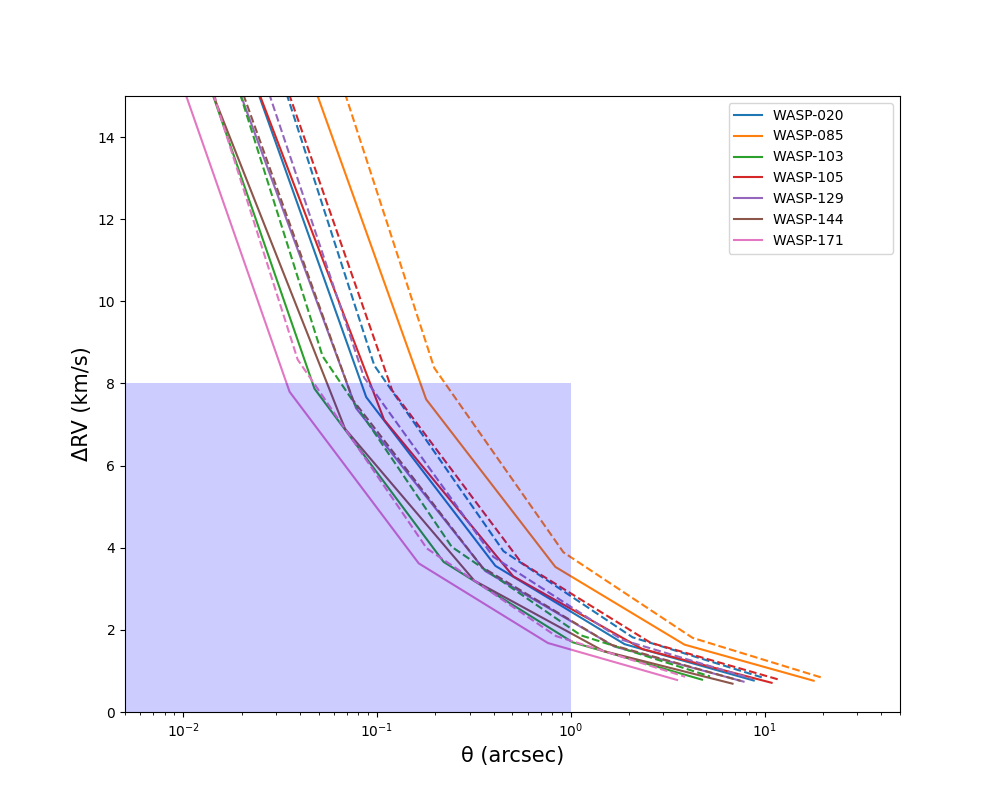}
    \caption{Radial velocity difference ($\Delta RV$ in km/s) vs angular separation ($\theta$) in arcsec. Each curve corresponds to 
the marked WASP-planet host, and each point on a curve corresponds to a particular value for orbital period. The portion of each curve that lies within the blue box corresponds to the parameters that the system should have in order to escape detection as a “stealth binary”. \textcolor{black}{Solid lines denote a model assuming that the total mass of the binary system is 1.5 times the primary mass. Dashed lines show the effect of assuming the total mass to be twice that of the primary.}}
    \label{fig:multicolor}
\end{figure}

In Fig.~\ref{fig:multicolor} we plot the $\Delta RV$ values obtained 
using Eq. \ref{eq:deltaRV} as a  function of the angular separation $\theta$ 
computed with Eq.~\ref{eq:theta} for
each of the seven WASP host stars with anomalous radii, on 
a logarithmic grid of periods spanning the range $1-10^{5}$ years.
The total mass is assumed to be \textcolor{black}{ between 1.5 and 2 times that of the planet-hosting star. Planets are preferentially identified around the brighter components of binaries when the observations are signal-to-noise-limited. The mass ratio distribution of binary systems is quite flat, so a reasonable assumption might be that the total mass is on average $\sim 1.5$ times that of the planet-host star if its stellar companion is bright enough to affect the overall angular diameter. If we also make allowance for cases where the luminosity ratio is close to 1 and the planet orbits the fainter star, an assumed mass ratio closer to 2 might be preferable. Figure 3 shows that the difference between these two assumptions has little practical impact on the inferred angular and radial-velocity separations.}

The blue box indicates the limits on $\Delta RV$ and $\theta$ 
below which we anticipate the binary to remain undetected; i.e., to lie within 
the hiding zone. Each point on the curve for any 
WASP host corresponds to a particular estimate on 
the period of the two stars. Thus, using the 
boundaries of our “stealth binaries box”, we 
calculate the minimum and maximum period, as 
well as the minimum possible angular separation. 
These are summarised in Table \ref{tab:Tab1}. \textcolor{black}{Increasing 
the assumed system mass to twice that of the primary decreases
the maximum period by about 15 percent and increases 
the minimum period by 30 percent. The minimum angular 
separation increases by about 35 percent.}

\begin{table}
	\centering
	\caption{Minimum and maximum orbital periods and 
angular separations ($\theta _{\text{max}} = 1$") corresponding to the 
stealth binaries box \textcolor{black}{assuming a system mass 1.5 times the primary mass.}}
	\label{tab:Tab1}
    \textcolor{black}{
	\begin{tabular} {|c|c|c|c|}
		\hline
		\textbf{WASP host star} & \textbf{$P_{\text{max}}$(yrs)} & \textbf{$P_{\text{min}}$(yrs)} & \textbf{$\theta _{\text{min}}$(arcsec)}\\
		\hline
		WASP-020 & 3837 	 & 88 & 0.08\\
		WASP-085 & 1319 	 & 86 & 0.16\\
		WASP-103 & 9651 	 & 95 & 0.05\\
        WASP-105 & 2792 	 & 70   & 0.09\\
        WASP-129 & 4600	 & 79 & 0.07\\
        WASP-144 & 5600 	 & 64   & 0.05\\
        WASP-171 & 15049 & 93 & 0.03\\
		\hline
	\end{tabular}
 }
\end{table}

The periods obtained are in the range of a hundred 
to a few thousand years. This explains naturally why none of these
systems could have been detected as a spectroscopic binary in 
the decade or so since their discovery.

\subsection{WASP-85}
\label{sec:w85}

\begin{figure}
	\includegraphics[width=\columnwidth]{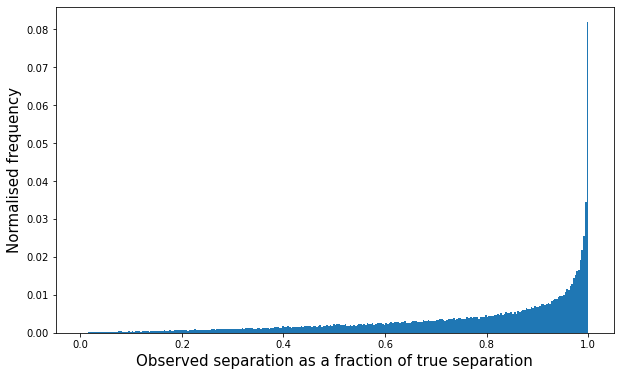}
    \caption{Probability distribution of the ratio of observed angular separation to the true angular separation. For an angle $\theta$ between the line connecting the two stars and the line of sight to the observer, $\sin{\theta}$ gives us this ratio. The curve has been obtained by sampling from a uniform distribution in $\cos{\theta}$ between -1 and 1 to calculate the corresponding values of $\sin{\theta}$ \textcolor{black}{\citep[cf.][]{2016AndrewCameron,Hatzes2016}.}}
    \label{fig:bluecurve}
\end{figure}

The binary companion of WASP-85A was detected 
during its discovery by \cite{2014arXiv1412.7761B}. From 
the discovery paper, we took the values for the mass 
of the companion, and the observed angular 
separation from {\em Gaia} DR3 to get the 
corresponding orbital period from Eq. \ref{eq:theta}. 
To estimate the true orbital separation from the observed angular separation, 
we must take into account the random orbit orientation. The distribution of apparent angular
separation as a fraction of true separation depends on the angle between the 
line connecting the stellar centres (assumed randomly oriented in space) and the line of sight. Fig. \ref{fig:bluecurve}
shows this probability distribution, with a mean of 
0.79. Thus, on average, we only observe 80\% of the 
true angular separation.

The {\em Gaia} DR3 catalogue gives the observed angular 
separation of the WASP-85AB binary system as 
$\theta= 1.28''$ ; this suggests a true angular 
separation of about $1.6''$. On the plot of $\Delta RV$ vs $\theta$ in Fig.~\ref{fig:multicolor}, 
this angular separation corresponds to a $\Delta RV$ of 2.78 km/s and an 
orbital period of ~2500 years, which is in good 
agreement with the estimate of 2000-3000 years in 
the discovery paper. 

The companion star (WASP-85B) is about 0.9 mag 
fainter in the G band than the primary star, with a 
flux ratio of about 0.5 in most bands. The discovery paper gives a mass 
$M_B=0.88 \pm  0.07 M_{\odot}$ and a radius $R_B=0.77 \pm  0.13 R_{\odot}$. 
If we take the photodynamical radius of the primary $R_{A,{\rm trans}}=0.935\,\textcolor{black}{R_\odot}$
and scale it by the square root of the flux ratio $f_{A+B}/f_A\simeq 1.5$, we
expect to obtain $R_{\rm IRFM}=1.16 R_{\odot}$ when the light of both components is combined.

Since Gaia was able to resolve the system, we combined the 
Gaia magnitudes for the primary and secondary stars to 
recalculate the total flux coming from the system 
and hence the combined Gaia  $BP$, $G$ and $RP$ magnitudes. 
Combining these with the
2MASS and WISE magnitudes (in which the binary is unresolved),
the resulting angular diameter and parallax yielded 
$R_{\rm IRFM}=1.16  R_{\odot}$. This confirms that the angular radius
derived from the combined spectral energy distribution of both
binary components is overestimated by an amount that is 
consistent with our knowledge of the two stars. This means that in cases 
where $R_{\rm trans} < R_{\text{IRFM}}\lesssim \sqrt{2}R_{\rm trans}$, there
is a strong possibility that the planet orbits the brighter component of a stealth 
binary. There may indeed be cases where the luminosity ratio is close to 1, and 
a planet orbiting the fainter component produces detectable transits. 
In such cases, it would be possible to obtain 
$R_{\text{IRFM}}\gtrsim \sqrt{2}R_{\rm trans}$. This could explain the location
of the system WASP-20, above the orange line in Fig.~\ref{fig:TEPCat}, as discussed in \textcolor{black}{Section~\ref{sec:specialcases}}.

\subsection{Predicting flux and mass ratios}

We now use the information we have about the 
companion star to make estimates on the flux and 
mass ratios with respect to the primary star for each 
of our outliers. We use isochrones from 
\cite{2012MNRAS.427..127B}
to relate stellar masses 
to magnitudes. Table \ref{tab:Tab2} shows some of the 
parameters of the primary stars that we have used 
for our analysis in this section. The stellar radii and 
masses have been taken from TEPCat, while the 
rough age estimates have been taken from the 
discovery papers.

\begin{table}
	\centering
	\caption{Some parameters of the outliers taken from 
TEPCat and discovery papers, used for further analysis 
in this section.}
	\label{tab:Tab2}
	\begin{tabular} {|c|c|c|c|c|c|c|}
		\hline
		\textbf{WASP host star} & \textbf{$R_{\text{S}}$ (in $R_{\odot}$)} & \textbf{$M_{\text{S}}$ (in $M_{\odot}$)} & \textbf{Age (Gyrs)}\\
		\hline
		WASP-020 & 1.242 & 1.113 & 4.0\\
		WASP-085 & 0.935 & 1.090 & 0.3\\
		WASP-103 & 1.413 & 1.205 & 4.0\\
        WASP-105 & 0.900 & 0.890 & 6.0\\
        WASP-129 & 0.900 & 1.000 & 1.0\\
        WASP-144 & 0.810 & 0.810 & 8.0\\
        WASP-171 & 1.637 & 1.171 & 6.0\\
		\hline
	\end{tabular}
\end{table}

\begin{figure}
	\includegraphics[width=\columnwidth]{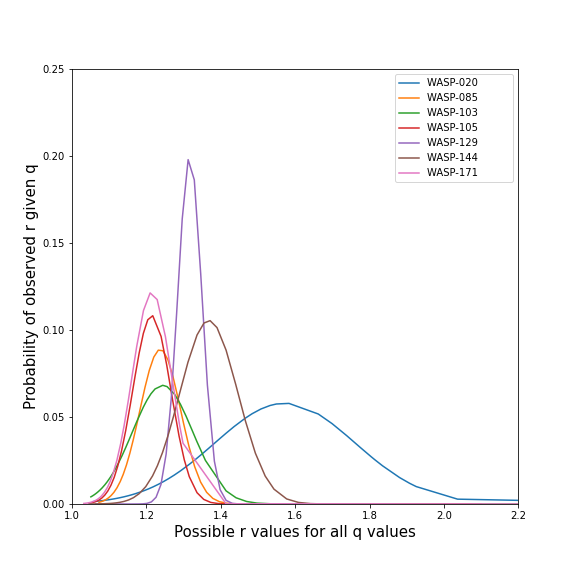}
    \caption{Probability curves for the possible values of the factor $r$ defined in Section 4.3, given the value that was obtained from the $R_{\text{IRFM}}$ calculation for each host.}
    \label{fig:Q}
\end{figure}

Using these isochrones, we estimated the ratio of fluxes of the primary to the companion star. We evaluated the flux ratio for every mass ratio $q = M_2/M_1$ from 0.7 to 1.2 in intervals of 0.01. We also computed the most probable flux ratio from the factor $r$ by which $R_{\text{IRFM}}$ is greater than the actual radius; i.e., $1+\frac{f_{2}}{f_{1}} = r^{2}$ where $f_{2}$ and $f_{1}$ are the fluxes received from the secondary and primary stars respectively. Each value of $r$ corresponds to a different value of $q$, found by interpolating the isochrone for a star of the same age, assuming that the age of the companion is the same as that of the primary star. These results are summarised in Table \ref{tab:Tab3}, while Fig. \ref{fig:Q} shows the probability curves for the factor $r$, given the value that was observed.

\begin{table*}
	\centering
	\caption{Most probable value for factor $r$, with mass ratio and flux ratio, and errors.}
	\label{tab:Tab3}
	\begin{tabular} {ccccc}
		\hline
		\textbf{WASP host star} & \textbf{Most Probable $r$} & \textbf{Observed $r$} & \textbf{Most Probable Flux Ratio} & \textbf{Most Probable Mass Ratio}	\\
		\hline
		WASP-020 & 1.58 $\pm$ 0.19 & 1.57 & 1.51 $\pm$ 0.36 & 1.07 \\
		WASP-085 & 1.23 $\pm$ 0.05 & 1.24 & 0.52 $\pm$ 0.05 & 0.88 \\
		WASP-103 & 1.24 $\pm$ 0.08 & 1.24 & 0.55 $\pm$ 0.07 & 0.92 \\
        WASP-105 & 1.22 $\pm$ 0.05 & 1.21 & 0.48 $\pm$ 0.04 & 0.89 \\
        WASP-129 & 1.31 $\pm$ 0.03 & 1.32 & 0.72 $\pm$ 0.04 & 0.94 \\
        WASP-144 & 1.37 $\pm$ 0.08 & 1.37 & 0.88 $\pm$ 0.10 & 0.98 \\
        WASP-171 & 1.21 $\pm$ 0.05 & 1.21 & 0.46 $\pm$ 0.04 & 0.94 \\
		\hline
	\end{tabular}
\end{table*}

\begin{table*}
    \centering
    \caption{Final corrected planetary parameters for all the outliers, after having accounted for the additional flux from the secondary star. The fractional error in each quantity remains the same. Original references for photodynamical values adopted from TEPCat are given as footnotes. Corrections have been made using the relations discussed in Section 4.4. 
    }
	\label{tab:Tab4}
    \resizebox{\textwidth}{!}{\begin{tabular}{ccccccc}
    \hline
    \textbf{WASP host star} & \textbf{Photodynamical $R_{p}$ (in $R_{J}$)} & \textbf{Corrected $R_{p}$ (in $R_{J}$)} & \textbf{Photodynamical $M_{P}$ (in $M_{J}$)} & \textbf{Corrected  $M_{P}$ (in  $M_{J}$)} & \textbf{Photodynamical $\rho_{P}$ (in $\rho_{J}$)} & \textbf{Corrected $\rho_{P}$ (in $\rho_{J}$)} \\
        \hline
        WASP-020$^a$ & 1.46 & 2.31 $\pm$ 0.29  & 0.31  & 0.78 $\pm$ 0.19  & 0.10 & 0.06 $\pm$ 0.01  \\
        WASP-085*  & 1.01 & \textit{1.25 $\pm$ 0.07} & 0.82  & \textit{1.25 $\pm$ 0.03} & 0.80  &     \textit{0.65 $\pm$ 0.02} \\
        WASP-103$^b$ & 1.53 & 1.90 $\pm$ 0.14 & 1.49 & 2.31 $\pm$ 0.33 & 0.42 & 0.34 $\pm$ 0.05  \\
        WASP-105$^c$ & 0.96 & 1.17 $\pm$ 0.06  & 1.80 & 2.67 $\pm$ 0.27 & 2.00 & 1.64   $\pm$ 0.11 \\
        WASP-129$^d$ & 0.93 & 1.22 $\pm$ 0.05 & 1.00 & 1.72 $\pm$ 0.19 & 1.20 & 0.91 $\pm$ 0.15 \\
        WASP-144$^e$ & 0.85 & 1.17  $\pm$ 0.10 & 0.44 & 0.83 $\pm$ 0.15 & 0.72 & 0.53 $\pm$ 0.11 \\
        WASP-171$^f$ & 0.98 & 1.19 $\pm$ 0.08 & 1.08 & 1.59 $\pm$ 0.20 & 1.13 & 0.93 $\pm$ 0.17 \\
        \hline
        \multicolumn{7}{l}{$^a$ Photodynamical values from \citet{And15}}\\
        \multicolumn{7}{l}{$^*$ Photodynamical values reverse-calculated using $r$ value from Table \ref{tab:Tab3},
        from values (in italics) already corrected for binarity in \citet{2014arXiv1412.7761B} (see Sect.~\ref{sec:w85})}\\
        \multicolumn{7}{l}{$^b$ Photodynamical values from \citet{2014A&A...562L...3G}}\\
        \multicolumn{7}{l}{$^c$ Photodynamical values from \citet{2017A&A...604A.110A}}\\
        \multicolumn{7}{l}{$^d$ Photodynamical values from \citet{2016A&A...591A..55M}}\\
        \multicolumn{7}{l}{$^e$ Photodynamical values from \citet{2019MNRAS.482.1379H}}\\
        \multicolumn{7}{l}{$^f$ Photodynamical values from \citet{2019MNRAS.489.2478N}}\\
    \end{tabular}}
\end{table*}

All the most-probable mass ratios are within the range 0.89-1, barring WASP-20. For this system,  r has a value greater than $\sqrt{2}$, implying that $f_{2}$ > $f_{1}$; i.e., the secondary star is brighter than the primary, and the planet is orbiting the fainter and hence denser star. The identity of the planet-hosting component of the binary is discussed in Section~\ref{sec:specialcases}.

\subsection{Recalculating planetary parameters}

Using the most probable flux ratios calculated in the previous section, we can obtain the value of the factor ‘$r$’ and use this to correct the planet’s radius $R_{p}$, mass $M_{P}$, and density $\rho_{P}$. We know that:
\begin{equation*}
	\sqrt{\delta} \propto R_{p} \text{ and } \delta\propto R^{2} \Rightarrow \boldsymbol{R_{p} \propto r}
\end{equation*}
where $\delta$ is the transit depth. From \cite{2016AndrewCameron}:
\begin{equation*}
	\rho_{P} \propto \frac{K}{(R_{p})^{3}}\textit{\rm\ and\ }K \propto r^{2} \Rightarrow \boldsymbol{\rho_{P}\propto r^{-1}\textit{\rm\ and\ } M_{P} \propto r^{2}}
\end{equation*}

$K$ is the RV amplitude, whose observed value is the flux-weighted average of all the $K$ values for each star. Using the above relations and the published values of these quantities, we recalculated the planetary parameters, as summarised in Table \ref{tab:Tab4}. 

The surface gravity of the planet varies directly with its mass and inversely with the square of its radius. After accounting for the stealth binarity; the mass goes up by $r^{2}$ while the radius goes up by $r$ – which means that the planet's surface gravity remains unchanged. This is an interesting result, as it tells us that the planetary surface gravity is immune to stealth binaries.

\subsection{WASP-20 and WASP-103}
\label{sec:specialcases}

The binary companion of WASP-20A was 
discovered at a separation of 0.26 arcsec by \cite{2016ApJ...833L..19E} using near-IR adaptive-optics imaging with the SPHERE instrument on the VLT. They inferred an 
increase in planetary mass $M_{P}$ and radius $R_{p}$ by $4\sigma$ 
and $1\sigma$ respectively, after having accounted for the 
dilution to the transit curve and radial velocity 
amplitude. They concluded that the planet orbits the brighter of the two stars, on the grounds that the inferred stellar density in the planet-orbits-fainter-star scenario yielded an implausibly old stellar age of 16 Gyr. The inferred separation of 61 au and system mass of order 2$M_\odot$ implies a period of order 340 years, and hence a radial-velocity separation $K_A+K_B\simeq 5.3$ km s$^{-1}$,
placing it within the blue box in Fig.~\ref{fig:multicolor}.

\cite{2020A&A...635A..74S} made further 
corrections to these parameters using updated light 
curves and spectroscopic data. They were unable to rule out the planet-orbits-fainter-star scenario, finding a density 1.08 times solar and an age of 3.3 Gyr. In both cases, the planet's density is reduced relative to that obtained by assuming that the host star is single. In our own study, the balance of probability in Fig. \ref{fig:Q} suggests that the ratio of the angular and photodynamical radii is greater than 1.4 for an assumed age of 4 Gyr. Our study thus favours the planet-orbits-fainter-star scenario. 

The planetary parameters for WASP-103b were also 
corrected in the analysis by \cite{2018MNRAS.474.2334D} for the presence 
of a stellar companion at a separation of 0.23 arcsec. This companion was
discovered in a lucky-imaging survey by \cite{2015A&A...579A.129W}. 
It is 3.1 magnitudes fainter than the brighter star in the $i'$ band 
and 2.6 mag fainter in the $z'$ band. We find that a discrepancy remains 
between $R_{\rm IRFM}$ and the value of $R_{\rm trans}$ found by \cite{2018MNRAS.474.2334D} even
after this correction, so we infer that the host star may have 
another brighter and closer companion. 

In their study of the tidal deformation and orbital decay rate of WASP-103b,
\cite{2022A&A...657A..52B} recently found that the 
the faint visual companion of the host star was too distant and insufficiently massive to explain the inferred positive RV acceleration $a=+0.113 \pm 0.058$ m~s$^{-1}$ day$^{-1}$, whose sign is contrary to expectations for tidal orbit decay. This suggests that the unresolved stellar companion responsible for the observed excess in the stellar angular diameter could also be the cause of the anomalous observed acceleration. Using our derived value $q=0.92$ for the stealth companion of WASP-103, we get $M_{2}=1.11 \pm 0.10 M_{\odot}$, where $M_{2}$ is the mass of the companion. If we assume that the unseen companion is close to superior conjunction, the host star should be accelerating away from the observer. Using the observed acceleration and the inferred secondary mass, assuming a circular orbit and a high inclination we obtain an approximate estimate of the orbital separation:
\begin{equation*}
    d=\sqrt{\frac{GM_{2}}{a}}
    \simeq 71 \pm 18\ {\rm au.}
\end{equation*}
where $d$ is the binary separation. At maximum elongation this would give an angular separation $\theta=0.15\pm0.04$ arcsec and an orbital period $P\simeq 400$ y. The maximum radial-velocity separation of the two stars would then be
$K_1+K_2=2\pi d/P \simeq 1.1$ au/yr, or 5.2 km s$^{-1}$. These inferred maximum angular and radial-velocity separations lie comfortably within the blue region in Fig.~\ref{fig:multicolor}, explaining why the inferred binary companion has thus far evaded detection.

\section{Conclusion and suggested work}

Various properties of transiting exoplanets can be derived from their transit photometry and RV measurements. If these get contaminated by other stars that are bound to the host star, then the calculated parameters need to be corrected by accounting for the additional flux in the system. We find that the surface gravity of the planet is a quantity that is not affected by the dilution of the transit and RV observations. We have recalculated the masses, radii, and densities for \textcolor{black}{seven} WASP planets (Table \ref{tab:Tab4}) where we suspect that the host star has a hidden stellar companion. 
Our estimate of the flux ratio and orbital period for the WASP-85 binary system is  consistent with that obtained in the discovery paper of \cite{2014arXiv1412.7761B}. Dilution of observations for WASP-20 and WASP-103 had been accounted for previously in literature.
Our comparisons of the $R_\text{IRFM}$ values with estimates of the host-star radius based on the photodynamical stellar density suggest that the planet in the WASP-20 system may orbit the fainter of the two stars, and that the stealth companion of WASP 103's host star may be responsible for the apparent secular increase in the planet's orbital period.

Overall, the densities of the planets have gone down on average by a factor of 1.3. We also made estimates on the orbital periods, mass ratios and flux ratios for all \textcolor{black}{seven} stealth binary systems, which have been outlined in Tables \ref{tab:Tab1} and \ref{tab:Tab3}. The most probable flux ratios range from about 0.5 to 1, and mass ratios from 0.89 to 1 - barring WASP-20, which is an exception wherein both ratios are above 1 implying that the planetary host star is the denser and fainter of the two stellar binary components. The orbital periods range from a hundred to a few thousand years, which explains why the secondary stars have not been detected in radial-velocity observations yet. Nonetheless, the example of WASP-103 discussed above suggests that long-term RV monitoring could reveal secular accelerations in systems with companions of unequal luminosity.

These binary systems appear to have angular separations below \textcolor{black}{1$^{\prime\prime}$} but are not close enough to give resolvable Doppler shifts. We also suggest that follow-up observations be made using speckle imaging, lucky imaging, or adaptive optics. Speckle imaging removes effects of turbulence in the atmosphere and provides simultaneous photometric and astrometric data at sub-arcsecond precisions \citep{2019AJ....157..211M}. The ‘Differential Speckle Survey Instrument’ \citep{2009AJ....137.5057H} has successfully detected binary companions to stars in the ‘Kilodegree Extremely Little Telescope’ survey \citep{2018AJ....155...27C}, and can thus be used for these WASP systems as well - as was done 
\textcolor{black}{for} WASP-103 \citep{2015A&A...579A.129W}.

\section*{Data Availability}

The stellar and planetary data for the WASP systems investigated in this research are available in the TEPCat database curated by Dr John Southworth at the University of Keele.
All other research data underpinning this publication and the PYTHON
code and notebook used to prepare all diagrams in this paper will
be made available through the University of St Andrews Research
Portal.
The research data supporting this publication can be accessed at https://doi.org/10.17630/1ceff0e3-c2aa-40d0-996a-0d3f8d81cf19 \citep{goswamy2024a}.
The code supporting this publication can be accessed at https://doi.org/10.17630/72a692c8-dd85-4078-99d3-8a5551727ff9 \citep{goswamy2024b}.

\section*{Acknowledgements}

Tanvi Goswamy thanks Siddharth Rangnekar for helpful discussions and assistance with preparation of the manuscript. 
Andrew Collier Cameron and Thomas Wilson acknowledge support from STFC consolidated grant numbers ST/R000824/1 and ST/V000861/1, and UKSA grant number ST/R003203/1.
This research has made use of NASA’s Astrophysics Data System.
This research has made use of the VizieR catalogue access tool, CDS, Strasbourg, France.



\bibliographystyle{mnras}



\onecolumn
\appendix
\section{}


\begin{longtable}{lll}
\caption{R\_IRFM and R\_trans values along with their errors, used in Fig. 1}
\label{tab:my-table}\\
\hline
\textbf{WASP host star} & \textbf{R from IRFM} & \textbf{Published R\_trans}\\
\hline
\endfirsthead
\multicolumn{3}{c}%
{{\bfseries Table \thetable\ continued from previous page}} \\
\hline
\textbf{WASP host star} & \textbf{R from IRFM} & \textbf{Published R\_trans} \\
\hline
\endhead
WASP-001                & 1.506 ± 0.015        & 1.470 ± 0.027        \\
WASP-002                & 0.884 ± 0.013        & 0.821 ± 0.014        \\
WASP-003                & 1.339 ± 0.011        & 1.298 ± 0.049        \\
WASP-004                & 0.912 ± 0.007        & 0.910 ± 0.018        \\
WASP-005                & 1.106 ± 0.008        & 1.088 ± 0.040        \\
WASP-006                & 0.817 ± 0.006        & 0.864 ± 0.025        \\
WASP-007                & 1.457 ± 0.009        & 1.478 ± 0.088        \\
WASP-008                & 0.994 ± 0.012        & 0.976 ± 0.021        \\
WASP-010                & 0.736 ± 0.011        & 0.678 ± 0.030        \\
WASP-011                & 0.858 ± 0.013        & 0.772 ± 0.015        \\
WASP-012                & 1.706 ± 0.025        & 1.657 ± 0.045        \\
WASP-013                & 1.544 ± 0.008        & 1.657 ± 0.079        \\
WASP-014                & 1.300 ± 0.011        & 1.318 ± 0.084        \\
WASP-015                & 1.420 ± 0.011        & 1.522 ± 0.044        \\
WASP-016                & 1.084 ± 0.008        & 1.087 ± 0.042        \\
WASP-017                & 1.579 ± 0.019        & 1.583 ± 0.041        \\
WASP-018                & 1.244 ± 0.007        & 1.255 ± 0.028        \\
WASP-019                & 1.003 ± 0.007        & 1.018 ± 0.015        \\
WASP-020                & 1.953 ± 0.223        & 1.242 ± 0.045        \\
WASP-021                & 1.178 ± 0.009        & 1.136 ± 0.051        \\
WASP-022                & 1.212 ± 0.017        & 1.255 ± 0.030        \\
WASP-023                & 0.824 ± 0.007        & 0.819 ± 0.031        \\
WASP-024                & 1.376 ± 0.013        & 1.317 ± 0.041        \\
WASP-025                & 0.916 ± 0.006        & 0.924 ± 0.018        \\
WASP-026                & 1.303 ± 0.011        & 1.284 ± 0.036        \\
WASP-028                & 1.107 ± 0.008        & 1.083 ± 0.025        \\
WASP-029                & 0.787 ± 0.011        & 0.808 ± 0.044        \\
WASP-030                & 1.440 ± 0.013        & 1.389 ± 0.029        \\
WASP-031                & 1.280 ± 0.012        & 1.252 ± 0.033        \\
WASP-032                & 1.155 ± 0.013        & 1.110 ± 0.050        \\
WASP-033                & 1.548 ± 0.016        & 1.509 ± 0.025        \\
WASP-034                & 1.057 ± 0.005        & 0.930 ± 0.120        \\
WASP-035                & 1.119 ± 0.008        & 1.090 ± 0.030        \\
WASP-036                & 0.943 ± 0.008        & 0.985 ± 0.014        \\
WASP-037                & 1.043 ± 0.012        & 1.003 ± 0.053        \\
WASP-038                & 1.495 ± 0.015        & 1.331 ± 0.028        \\
WASP-039                & 0.925 ± 0.006        & 0.939 ± 0.022        \\
WASP-041                & 0.892 ± 0.005        & 0.886 ± 0.012        \\
WASP-042                & 0.862 ± 0.006        & 0.892 ± 0.021        \\
WASP-043                & 0.678 ± 0.018        & 0.667 ± 0.011        \\
WASP-044                & 0.936 ± 0.012        & 0.865 ± 0.038        \\
WASP-045                & 0.908 ± 0.012        & 0.917 ± 0.024        \\
WASP-046                & 0.908 ± 0.008        & 0.858 ± 0.027        \\
WASP-047b               & 1.151 ± 0.010        & 1.137 ± 0.013        \\
WASP-047c               & 1.150 ± 0.010        & 1.137 ± 0.013        \\
WASP-047d               & 1.151 ± 0.011        & 1.137 ± 0.013        \\
WASP-048                & 1.757 ± 0.014        & 1.519 ± 0.051        \\
WASP-049                & 1.024 ± 0.011        & 1.038 ± 0.037        \\
WASP-050                & 0.876 ± 0.005        & 0.855 ± 0.019        \\
WASP-052                & 0.841 ± 0.008        & 0.786 ± 0.016        \\
WASP-053                & 0.837 ± 0.006        & 0.798 ± 0.023        \\
WASP-054                & 1.699 ± 0.012        & 1.828 ± 0.086        \\
WASP-055                & 1.087 ± 0.009        & 1.102 ± 0.019        \\
WASP-056                & 1.199 ± 0.010        & 1.112 ± 0.024        \\
WASP-057                & 1.079 ± 0.009        & 0.927 ± 0.033        \\
WASP-058                & 1.172 ± 0.011        & 1.170 ± 0.130        \\
WASP-059                & 0.710 ± 0.015        & 0.613 ± 0.044        \\
WASP-060                & 1.488 ± 0.015        & 1.401 ± 0.066        \\
WASP-061                & 1.356 ± 0.010        & 1.390 ± 0.030        \\
WASP-062                & 1.238 ± 0.008        & 1.280 ± 0.050        \\
WASP-063                & 1.755 ± 0.013        & 1.880 ± 0.080        \\
WASP-064                & 1.072 ± 0.011        & 1.058 ± 0.025        \\
WASP-065                & 1.101 ± 0.009        & 1.010 ± 0.050        \\
WASP-066                & 1.785 ± 0.037        & 1.750 ± 0.090        \\
WASP-067                & 0.867 ± 0.006        & 0.817 ± 0.022        \\
WASP-068                & 1.684 ± 0.011        & 1.690 ± 0.085        \\
WASP-069                & 0.847 ± 0.019        & 0.813 ± 0.028        \\
WASP-070                & 1.250 ± 0.019        & 1.251 ± 0.079        \\
WASP-071                & 2.144 ± 0.021        & 2.260 ± 0.170        \\
WASP-072                & 2.180 ± 0.030        & 1.980 ± 0.240        \\
WASP-073                & 2.213 ± 0.017        & 2.070 ± 0.135        \\
WASP-074                & 1.531 ± 0.008        & 1.536 ± 0.026        \\
WASP-075                & 1.306 ± 0.012        & 1.270 ± 0.020        \\
WASP-076                & 1.836 ± 0.037        & 1.765 ± 0.071        \\
WASP-077                & 1.071 ± 0.028        & 0.955 ± 0.015        \\
WASP-078                & 2.031 ± 0.023        & 2.350 ± 0.095        \\
WASP-079                & 1.590 ± 0.010        & 1.510 ± 0.035        \\
WASP-080                & 0.650 ± 0.015        & 0.593 ± 0.012        \\
WASP-081                & 1.241 ± 0.013        & 1.283 ± 0.039        \\
WASP-082                & 2.086 ± 0.015        & 2.219 ± 0.087        \\
WASP-083                & 1.057 ± 0.010        & 1.050 ± 0.050        \\
WASP-084                & 0.817 ± 0.006        & 0.768 ± 0.019        \\
WASP-085                & 1.157 ± 0.043        & 0.935 ± 0.023        \\
WASP-086                & 2.136 ± 0.016        & 1.291 ± 0.014        \\
WASP-087                & 1.623 ± 0.019        & 1.627 ± 0.062        \\
WASP-088                & 2.131 ± 0.029        & 2.080 ± 0.090        \\
WASP-089                & 0.904 ± 0.010        & 0.880 ± 0.030        \\
WASP-090                & 1.892 ± 0.031        & 1.980 ± 0.090        \\
WASP-091                & 0.856 ± 0.007        & 0.860 ± 0.030        \\
WASP-092                & 1.286 ± 0.016        & 1.341 ± 0.058        \\
WASP-093                & 1.629 ± 0.019        & 1.524 ± 0.040        \\
WASP-094                & 1.567 ± 0.012        & 1.620 ± 0.045        \\
WASP-095                & 1.213 ± 0.010        & 1.130 ± 0.060        \\
WASP-096                & 1.081 ± 0.010        & 1.050 ± 0.050        \\
WASP-097                & 1.090 ± 0.008        & 1.060 ± 0.040        \\
WASP-098                & 0.735 ± 0.006        & 0.741 ± 0.021        \\
WASP-099                & 1.688 ± 0.013        & 1.760 ± 0.085        \\
WASP-100                & 1.723 ± 0.014        & 1.670 ± 0.145        \\
WASP-101                & 1.311 ± 0.009        & 1.290 ± 0.040        \\
WASP-102                & 1.389 ± 0.016        & 1.331 ± 0.013        \\
WASP-103                & 1.757 ± 0.100        & 1.413 ± 0.045        \\
WASP-104                & 0.935 ± 0.006        & 0.935 ± 0.010        \\
WASP-105                & 1.093 ± 0.028        & 0.900 ± 0.030        \\
WASP-106                & 1.481 ± 0.013        & 1.393 ± 0.038        \\
WASP-107                & 0.661 ± 0.014        & 0.670 ± 0.020        \\
WASP-108                & 1.247 ± 0.016        & 1.215 ± 0.040        \\
WASP-109                & 1.412 ± 0.019        & 1.346 ± 0.044        \\
WASP-110                & 0.876 ± 0.007        & 0.881 ± 0.035        \\
WASP-111                & 1.887 ± 0.015        & 1.850 ± 0.100        \\
WASP-112                & 1.082 ± 0.012        & 1.002 ± 0.037        \\
WASP-113                & 1.741 ± 0.015        & 1.608 ± 0.105        \\
WASP-114                & 1.414 ± 0.017        & 1.430 ± 0.060        \\
WASP-117                & 1.213 ± 0.007        & 1.170 ± 0.063        \\
WASP-118                & 1.858 ± 0.021        & 1.754 ± 0.016        \\
WASP-119                & 1.107 ± 0.009        & 1.200 ± 0.100        \\
WASP-120                & 1.707 ± 0.015        & 1.870 ± 0.110        \\
WASP-121                & 1.473 ± 0.008        & 1.440 ± 0.030        \\
WASP-122                & 1.454 ± 0.010        & 1.520 ± 0.030        \\
WASP-123                & 1.221 ± 0.017        & 1.285 ± 0.051        \\
WASP-124                & 1.147 ± 0.010        & 1.020 ± 0.020        \\
WASP-126                & 1.192 ± 0.013        & 1.270 ± 0.075        \\
WASP-127                & 1.354 ± 0.012        & 1.333 ± 0.027        \\
WASP-128                & 1.203 ± 0.010        & 1.152 ± 0.019        \\
WASP-129                & 1.185 ± 0.012        & 0.900 ± 0.020        \\
WASP-130                & 1.035 ± 0.008        & 0.960 ± 0.030        \\
WASP-131                & 1.673 ± 0.021        & 1.526 ± 0.065        \\
WASP-132                & 0.740 ± 0.010        & 0.740 ± 0.020        \\
WASP-133                & 1.507 ± 0.017        & 1.440 ± 0.050        \\
WASP-134                & 1.165 ± 0.010        & 1.175 ± 0.048        \\
WASP-135                & 0.889 ± 0.008        & 0.960 ± 0.050        \\
WASP-136                & 2.071 ± 0.019        & 2.210 ± 0.220        \\
WASP-137                & 1.634 ± 0.014        & 1.520 ± 0.110        \\
WASP-138                & 1.448 ± 0.013        & 1.360 ± 0.050        \\
WASP-139                & 0.818 ± 0.007        & 0.800 ± 0.040        \\
WASP-140                & 0.817 ± 0.011        & 0.870 ± 0.040        \\
WASP-141                & 1.316 ± 0.013        & 1.370 ± 0.070        \\
WASP-142                & 1.642 ± 0.022        & 1.640 ± 0.080        \\
WASP-143                & 0.992 ± 0.010        & 1.013 ± 0.032        \\
WASP-144                & 1.108 ± 0.032        & 0.810 ± 0.040        \\
WASP-145                & 0.674 ± 0.010        & 0.680 ± 0.070        \\
WASP-146                & 1.314 ± 0.020        & 1.232 ± 0.072        \\
WASP-147                & 1.441 ± 0.016        & 1.370 ± 0.080        \\
WASP-148                & 0.926 ± 0.008        & 1.030 ± 0.200        \\
WASP-150                & 1.710 ± 0.024        & 1.651 ± 0.027        \\
WASP-151                & 1.220 ± 0.011        & 1.181 ± 0.020        \\
WASP-153                & 1.654 ± 0.020        & 1.730 ± 0.095        \\
WASP-156                & 0.824 ± 0.005        & 0.760 ± 0.030        \\
WASP-157                & 1.091 ± 0.014        & 1.134 ± 0.051        \\
WASP-158                & 1.571 ± 0.025        & 1.390 ± 0.180        \\
WASP-159                & 2.065 ± 0.022        & 2.110 ± 0.100        \\
WASP-160                & 0.847 ± 0.008        & 0.872 ± 0.030        \\
WASP-161                & 1.612 ± 0.278        & 1.712 ± 0.078        \\
WASP-162                & 1.168 ± 0.010        & 1.110 ± 0.050        \\
WASP-163                & 1.194 ± 0.014        & 1.015 ± 0.039        \\
WASP-164                & 0.957 ± 0.012        & 0.932 ± 0.029        \\
WASP-165                & 1.699 ± 0.028        & 1.650 ± 0.090        \\
WASP-166                & 1.240 ± 0.010        & 1.220 ± 0.060        \\
WASP-167                & 1.916 ± 0.028        & 1.790 ± 0.050        \\
WASP-168                & 1.087 ± 0.007        & 1.120 ± 0.060        \\
WASP-169                & 2.247 ± 0.028        & 2.011 ± 0.139        \\
WASP-170                & 1.016 ± 0.017        & 0.938 ± 0.059        \\
WASP-171                & 1.989 ± 0.022        & 1.637 ± 0.069        \\
WASP-172                & 2.101 ± 0.031        & 1.910 ± 0.100        \\
WASP-173                & 1.069 ± 0.010        & 1.110 ± 0.050        \\
WASP-174                & 1.381 ± 0.014        & 1.347 ± 0.018        \\
WASP-175                & 1.244 ± 0.012        & 1.204 ± 0.064        \\
WASP-176                & 2.005 ± 0.022        & 1.925 ± 0.046        \\
WASP-177                & 0.802 ± 0.008        & 0.885 ± 0.046        \\
WASP-178                & 1.700 ± 0.018        & 1.670 ± 0.070        \\
WASP-180                & 1.177 ± 0.012        & 1.190 ± 0.060        \\
WASP-181                & 1.017 ± 0.010        & 0.965 ± 0.043        \\
WASP-182                & 1.241 ± 0.011        & 1.340 ± 0.030        \\
WASP-183                & 0.873 ± 0.008        & 0.871 ± 0.038        \\
WASP-184                & 1.761 ± 0.020        & 1.650 ± 0.090        \\
WASP-185                & 1.586 ± 0.015        & 1.500 ± 0.080        \\
WASP-189                & 2.343 ± 0.028        & 2.360 ± 0.030        \\
WASP-190                & 1.811 ± 0.028        & 1.600 ± 0.100        \\
WASP-192                & 1.317 ± 0.015        & 1.320 ± 0.070        \\
\hline
\end{longtable}


\bsp	
\label{lastpage}
\end{document}